\documentstyle[twoside]{article}

\catcode`\@=11 \long\def\@makefntext#1{ \protect\noindent \hbox to 3.2pt {\hskip-.9pt
$^{{\eightrm\@thefnmark}}$\hfil}#1\hfill}               

\def\@makefnmark{\hbox to 0pt{$^{\@thefnmark}$\hss}}    

\def\ps@myheadings{\let\@mkboth\@gobbletwo
\def\@oddhead{\hbox{}
\rightmark\hfil\eightrm\thepage}
\def\@oddfoot{}\def\@evenhead{\eightrm\thepage\hfil
\leftmark\hbox{}}\def\@evenfoot{}
\def\sectionmark##1{}\def\subsectionmark##1{}}

\oddsidemargin=\evensidemargin  

\newcounter{sectionc}\newcounter{subsectionc}\newcounter{subsubsectionc}
\renewcommand{\section}[1] {\vspace{12pt}\addtocounter{sectionc}{1}
\setcounter{subsectionc}{0}\setcounter{subsubsectionc}{0}\noindent
        {\tenbf\thesectionc. #1}\par\vspace{5pt}}
\renewcommand{\subsection}[1] {\vspace{12pt}\addtocounter{subsectionc}{1}
        \setcounter{subsubsectionc}{0}\noindent
        {\bf\thesectionc.\thesubsectionc. {\kern1pt \bfit #1}}\par\vspace{5pt}}
\renewcommand{\subsubsection}[1] {\vspace{12pt}\addtocounter{subsubsectionc}{1}
        \noindent{\tenrm\thesectionc.\thesubsectionc.\thesubsubsectionc.
        {\kern1pt \tenit #1}}\par\vspace{5pt}}

\newcounter{appendixc}
\newcounter{subappendixc}[appendixc]
\newcounter{subsubappendixc}[subappendixc]
\renewcommand{\thesubappendixc}{\Alph{appendixc}.\arabic{subappendixc}}
\renewcommand{\thesubsubappendixc}
        {\Alph{appendixc}.\arabic{subappendixc}.\arabic{subsubappendixc}}

\renewcommand{\appendix}[1] {\vspace{12pt}
        \refstepcounter{appendixc}
        \setcounter{figure}{0}
        \setcounter{table}{0}
        \setcounter{lemma}{0}
        \setcounter{theorem}{0}
        \setcounter{corollary}{0}
        \setcounter{definition}{0}
        \setcounter{equation}{0}
        \renewcommand{\thefigure}{\Alph{appendixc}.\arabic{figure}}
        \renewcommand{\thetable}{\Alph{appendixc}.\arabic{table}}
        \renewcommand{\theappendixc}{\Alph{appendixc}}
        \renewcommand{\thelemma}{\Alph{appendixc}.\arabic{lemma}}
        \renewcommand{\thetheorem}{\Alph{appendixc}.\arabic{theorem}}
        \renewcommand{\thedefinition}{\Alph{appendixc}.\arabic{definition}}
        \renewcommand{\thecorollary}{\Alph{appendixc}.\arabic{corollary}}
        \renewcommand{\theequation}{\Alph{appendixc}.\arabic{equation}}
        \noindent{\tenbf Appendix \theappendixc #1}\par\vspace{5pt}}
\newcommand{\subappendix}[1] {\vspace{12pt}
        \refstepcounter{subappendixc}
        \noindent{\bf Appendix \thesubappendixc. {\kern1pt \bfit #1}}
        \par\vspace{5pt}}
\newcommand{\subsubappendix}[1] {\vspace{12pt}
        \refstepcounter{subsubappendixc}
        \noindent{\rm Appendix \thesubsubappendixc. {\kern1pt \tenit #1}}
        \par\vspace{5pt}}

\newcommand{\textlineskip}{\baselineskip=13pt}
\newcommand{\smalllineskip}{\baselineskip=10pt}

\def\eightcirc{
\begin{picture}(0,0)
\put(4.4,1.8){\circle{6.5}}
\end{picture}}
\def\eightcopyright{\eightcirc\kern2.7pt\hbox{\eightrm c}}

\newcommand{\copyrightheading}[1]
        {\vspace*{-2.5cm}\smalllineskip{\flushleft
        {\footnotesize International Journal of Modern Physics E, #1}\\
        {\footnotesize $\eightcopyright$\, World Scientific Publishing
         Company}\\
         }}

\newcommand{\publisher}[2]{{\begin{center}\footnotesize\smalllineskip
        Received #1\\
        Revised #2
        \end{center}
        }}

\renewenvironment{thebibliography}[1]
        {\frenchspacing
         \ninerm\baselineskip=11pt
         \begin{list}{\arabic{enumi}.}
        {\usecounter{enumi}\setlength{\parsep}{0pt}
         \setlength{\leftmargin 12.7pt}{\rightmargin 0pt} 
         \setlength{\itemsep}{0pt} \settowidth
        {\labelwidth}{#1.}\sloppy}}{\end{list}}

\newcounter{itemlistc}
\newcounter{romanlistc}
\newcounter{alphlistc}
\newcounter{arabiclistc}

\newcommand{\fcaption}[1]{
        \refstepcounter{figure}
        \setbox\@tempboxa = \hbox{\footnotesize Fig.~\thefigure. #1}
        \ifdim \wd\@tempboxa > 5in
           {\begin{center}
        \parbox{5in}{\footnotesize\smalllineskip Fig.~\thefigure. #1}
            \end{center}}
        \else
             {\begin{center}
             {\footnotesize Fig.~\thefigure. #1}
              \end{center}}
        \fi}

\newcommand{\tcaption}[1]{
        \refstepcounter{table}
        \setbox\@tempboxa = \hbox{\footnotesize Table~\thetable. #1}
        \ifdim \wd\@tempboxa > 5in
           {\begin{center}
        \parbox{5in}{\footnotesize\smalllineskip Table~\thetable. #1}
            \end{center}}
        \else
             {\begin{center}
             {\footnotesize Table~\thetable. #1}
              \end{center}}
        \fi}

\def\@citex[#1]#2{\if@filesw\immediate\write\@auxout
        {\string\citation{#2}}\fi
\def\@citea{}\@cite{\@for\@citeb:=#2\do
        {\@citea\def\@citea{,}\@ifundefined
        {b@\@citeb}{{\bf ?}\@warning
        {Citation `\@citeb' on page \thepage \space undefined}}
        {\csname b@\@citeb\endcsname}}}{#1}}

\newif\if@cghi
\def\cite{\@cghitrue\@ifnextchar [{\@tempswatrue
        \@citex}{\@tempswafalse\@citex[]}}
\def\citelow{\@cghifalse\@ifnextchar [{\@tempswatrue
        \@citex}{\@tempswafalse\@citex[]}}
\def\@cite#1#2{{$\null^{#1}$\if@tempswa\typeout
        {IJCGA warning: optional citation argument
        ignored: `#2'} \fi}}

\def\pmb#1{\setbox0=\hbox{#1}
        \kern-.025em\copy0\kern-\wd0
        \kern.05em\copy0\kern-\wd0
        \kern-.025em\raise.0433em\box0}

\def\fnt#1#2{\footnotetext{\kern-.3em
        {$^{\mbox{\scriptsize #1}}$}{#2}}}

\def\fpage#1{\begingroup
\voffset=.3in \thispagestyle{empty}\begin{table}[b]\centerline{\footnotesize #1}
        \end{table}\endgroup}

\def\runninghead#1#2{\pagestyle{myheadings}
\markboth{{\protect\footnotesize\it{\quad #1}}\hfill} {\hfill{\protect\footnotesize\it{#2\quad}}}} \headsep=15pt

\font\tenrm=cmr10 \font\tenit=cmti10

\font\tenbf=cmbx10 \font\bfit=cmbxti10 at 10pt \font\ninerm=cmr9  
\font\eightrm=cmr8    

\def\qed{\hbox{${\vcenter{\vbox{                        
   \hrule height 0.4pt\hbox{\vrule width 0.4pt height 6pt

   \kern5pt\vrule width 0.4pt}\hrule height 0.4pt}}}$}}

\def\bsc{{\sc a\kern-6.4pt\sc a\kern-6.4pt\sc a}}       
\def\bflatex{\bf L\kern-.30em\raise.3ex\hbox{\bsc}\kern-.14em
T\kern-.1667em\lower.7ex\hbox{E}\kern-.125em X}

\begin{document}

\runninghead{G.F. Marranghello, C.A.Z. Vasconcellos, M. Dillig and J. A. de Freitas Pacheco
} {A Nuclear Many-Body Theory at Finite Temperature Applied to  Protoneutron Star.
}

\normalsize\textlineskip \thispagestyle{empty} \setcounter{page}{1}

\copyrightheading{}                     

\vspace*{0.88truein}

\fpage{1} \centerline{\bf A Nuclear Many-Body Theory at Finite Temperature} \centerline{\bf Applied to  Protoneutron
Star.} \ \centerline{\bf } \vspace*{0.035truein}

\vspace*{0.015truein} \centerline{\footnotesize GUILHERME F. MARRANGHELLO\footnote{E-mail: gfm@if.ufrgs.br}, CESAR
A. Z. VASCONCELLOS
} \vspace*{0.015truein} \centerline{\footnotesize\it Instituto de F\'{\i}sica, Universidade Federal do Rio Grande do
Sul} \baselineskip=10pt \centerline{\footnotesize\it 91501-970 Porto Alegre - RS, Brasil
} \vspace*{10pt} \centerline{\footnotesize } \vspace*{10pt} \centerline{\footnotesize MANFRED DILLIG}
\vspace*{0.015truein} \centerline{\footnotesize\it Institut f\"ur Theoretische Physik III, der Universit\"at
Erlangen-N\"urnberg  } \baselineskip=10pt \centerline{\footnotesize\it
 D-91058 Erlangen, Germany }
 \vspace*{10pt}
 \centerline{\footnotesize and} \vspace*{10pt} \centerline{\footnotesize J. A. DE FREITAS PACHECO}
\vspace*{0.015truein} \centerline{\footnotesize\it Observatoire de la C\^ote d'Azur} \baselineskip=10pt
\centerline{\footnotesize\it
 Nice, France}
 \vspace*{0.225truein} \publisher{(received
date)}{(revised date)}

\vspace{0.5cm}
\begin{abstract}{ Thermodynamical properties of nuclear matter
are studied in the framework of an effective many-body field theory at finite temperature, considering the
Sommerfeld approximation. We perform the calculations by using the nonlinear Boguta and Bodmer model, extended by
the inclusion of the fundamental baryon octet and leptonic degrees of freedom. Trapped neutrinos are also included
in order to describe protoneutron star properties through the integration of the Tolman-Oppenheimer-Volkoff
equations, from which we obtain, beyond the standard relations for the masses and radii of protoneutron stars as
functions of the central density, new results of these quantities as functions of temperature. Our predictions
include: the determination of an absolute value for the limiting mass of protoneutron stars; new structural aspects
on the nuclear matter phase transition via the behaviour of the specific heat and, through the inclusion of quark
degrees of freedom, the properties of a hadron-quark phase transition and hybrid protoneutron stars.}{}{}
\end{abstract}

\section{ Introduction}

\indent

In the last few  decades, studies on the evolution of compact stars became a
central issue related to theoretical and
experimental research on the equation of state (EOS) of dense matter. In 1967, the first pulsar was observed\cite{hewish68} and, based on characteristic
observational features, this object was identified as a rotating neutron
star. Since this direct evidence for the existence of neutron stars,
nuclear models have been widely employed in the description of the internal structure, composition, dynamics and
evolution of these massive compact stars. As under the pull of gravity, the energy density in the core of compact
stars is thought to approach or even exceed the critical value of 1 $GeV/fm^3$, i.e. more than 6 times the density
of nuclear matter, their structure depends sensitively on the equation of state for very dense matter. As the

EOS describes how the energy density and pressure vary with density and temperature,

it should be also able to describe different phases, including gaseous nuclear
matter and liquid nuclei structure up to the deconfinement transition.

The EOS provided by model calculations and the Tolman-Oppenheimer-Volkoff (TOV)
equations\cite{tolman39,oppenheimer39} characterize the  structure of the star, whose  properties like
mass, radius, crust extent, moment of inertia are then susceptible to be
compared to observations\cite{glendenning97}. Moreover, these models should also provide the
description of relevant dynamical properties of compact stars as the rotational period, the emission of neutrinos
and gravitational waves as well as more fundamental aspects involving the evolution, internal composition and
structure of these stellar objects.

Concerning neutron stars, nuclear matter models should be able to account for such objects at least as massive as
the most massive observed pulsar, representing an important constraint on any proposed EOS for dense matter.
Presently, the best mass determination corresponds to the binary pulsar PSR 1913+16 ($M=1.444M_\odot$). Other
estimates based on quasi-periodic oscillations (QPOs), observed in the X-ray emission of low mass binaries (LMXBs)
suggest values up to $2.0-2.2M_\odot$, but these results are quite uncertain since they are model-dependent. More
recently, from new radial velocity data, the mass of the neutron star associated to the X-ray source in the system
Vela X-1\cite{pulsar} was recalculated with a larger accuracy than precedent estimates: the resulting value is
$1.86\pm 0.16M_\odot$. For a given family of stars, stiffer EOS predict higher limiting mass and larger radii ($ R
> 10-12$  km). In particular, recent hadrodynamics models point to EOS stiffer than those expected on the basis of a
non-relativistic approach, in better agreement with Vela X-1 results.

In the theoretical treatment of properties of compact stars, the relativistic phenomenological approach developed by
 Walecka\cite{walecka86},  quantum hadrodynamics (QHD), represents
one of the most important approaches to the highly nonlinear behavior of QCD at the hadronic energy scales. QHD
model is a relativistic quantum field theory based on a local lagrangian
 density which uses the nucleon and two
meson fields, the scalar-isoscalar, attractive $\sigma$ and the vector-isoscalar, repulsive $\omega$, as the
relevant effective degrees of freedom. This model provides a thermodynamically  consistent theoretical framework for
the description of bulk static properties of strong interacting many-body nuclear systems under extreme conditions.
The nonlinear model of Boguta and Bodmer (BB)\cite{boguta77} was developed to improve the description of  static
properties of nuclear matter, relative to the original Walecka model. With the introduction of  cubic and quartic
scalar meson self-interactions in the lagrangian density, the model has two additional parameters which allow and
provide sufficient flexibility to reproduce at saturation density, currently accepted values for  the compression
modulus of symmetric nuclear matter and the nucleon effective mass.

In this work we investigate static properties of nuclear matter at the high density domain by using a description
based on the BB model. As in its original form, the model is not yet suitable for an extrapolation to higher
densities, we considerably extent it by considering, at finite temperature, the fundamental baryon octet, lepton
degrees of freedom and trapped neutrinos in a consistent way, i.e. through the inclusion of the $\varrho$-meson and
the additional consideration of the equations for chemical equilibrium and charge neutrality. In section 3 we exceed
the limit of hadron degrees of freedom by including quarks and gluons in order
to allow a possible transition to quark matter. Within
this objective, we focus on a particular process in the evolution of compact
stars, preceding the supernova
explosion: the formation of a hot collapsed core or a {\it protoneutron star}\cite{prakash97}, which can reach
temperatures as high as few tens of MeV. These hot and dense relativistic objects are formed in a type-II supernova
explosion and evolves to a cold neutron star, basically through neutrino emission. Here, this very dense and hot
core is able to trap neutrinos, imparting momentum to the outer layers and then
cooling as it reaches a quasi-equilibrium state.

Hitherto, practically all microscopic investigations of stars have been performed for dense, cold matter in its
ground state. In this work, the evolution of protoneutron stars is studied,
extending previous investigations to non vanishing temperatures. The main novel
feature of the present study is the formulation of the dynamical equations in
the relativistic framework using a mean-field approach.  Global static properties as masses and radii are then computed as a function of central densities and
temperatures.

\section{ Nuclear Matter}

 Nowadays, astrophysical observations of compact stars and terrestrial nuclear physics experiments are
working together in the determination of the equation of state for dense matter. This allows to test the predictions
and consistency of different models for the properties of nuclear matter under extreme conditions, through the
comparison of both astrophysical data and results from high energy heavy ion collisions, where nuclear matter
approaches similar conditions of density and temperature as in the interior of dense compact stellar objects as
neutron and protoneutron stars.  Conversely, experiments with relativistic heavy ions at current facilities are now
providing first empirical information on the equation of state at high baryon matter densities, particularly
relevant to compact stars. Moreover, one feature is of particular interest: in addition to the deconfinement regime
at very high densities and temperatures up to $T_c=150MeV$ and beyond, there is another phase transition in nuclear
matter at much lower excitation energies, analogous to the liquid-gas transition in condensed matter physics. This
transition is predicted to occur with temperatures at $T_c>15MeV$ and hence is of particular interest for the
physics of supernova explosions, and in the formation of protoneutron stars. Intermediate energy heavy ion
collisions, at energies around 100 MeV per nucleon, are currently the most promising tools in searching for the
liquid-gas transition in laboratory.

The lagrangian density of our approach, which  takes into account the fundamental baryon octet, meson sectors, and
includes lepton degrees of freedom, is given by
\begin{eqnarray}
{\cal L}    &=& \sum\limits_{B}   \bar{\psi}_{B} [( i\gamma_\mu (\partial^\mu- g_{\omega B} \omega^{\mu}) -
(M_B-g_{\sigma B} \sigma)
]\psi_B \nonumber \\ && - \sum\limits_{B}   \bar{\psi}_{B} [
\frac12 g_{\varrho B} \mbox{\boldmath$\tau$} \cdot \mbox{\boldmath$\varrho$}^\mu] \psi_B
+\frac{bM}{3}\sigma^3+\frac{c}{4}\sigma^4 \nonumber \\ &&+\frac12(\partial_\mu \sigma \partial^\mu \sigma   -
{m_\sigma^2} \sigma^2)  - \frac14  \omega_{\mu \nu}  \omega^{\mu \nu}   + \frac12 {m_\omega^2}
 \omega_\mu \omega^\mu  \nonumber \\
&& -   \frac14 \mbox{\boldmath$\varrho$}_{\mu \nu} \cdot \mbox{\boldmath$\varrho$}^{\mu \nu} +  \frac12m_\varrho^2
\mbox{\boldmath$\varrho$}_\mu
 \cdot  \mbox{\boldmath$\varrho$}^\mu+
\sum\limits_{l}   \bar{\psi}_{l} [i \gamma_\mu \partial^\mu   - M_l] \psi_l \,\, .
\end{eqnarray}
This lagrangian density describes a system of eight baryons ($B = $ $p$, $n$, $\Lambda$, $\Sigma^-$, $\Sigma^0$,
$\Sigma^+$, $\Xi^-$, $\Xi^0$) coupled to three mesons ($\sigma$, $\omega$, $\varrho$) and leptons. The scalar and
vector coupling constants in the theory, $g_{\sigma}$ and $g_{\omega}$, and the coefficients $b$ and $c$ for the
nonlinear $\sigma$ self-couplings are determined to reproduce, at saturation density $\rho_0 = 0.153 fm^{-3}$, the
binding energy, $B = -16.3 MeV$, and the compression modulus, $K=240 MeV$ of nuclear matter, and the nucleon
effective mass, $M^\star=732MeV$. Additionally, the isovector coupling constant $g_\varrho$ is determined from the
symmetry energy coefficient, $a_4 = 32.5 MeV$, in nuclear matter. The hyperon/nucleon coupling constant ratios
$\chi_i=g_{Hi}/g_i$, with $i = \sigma, \omega$, are constrained through the binding energy of the $\Lambda$-hyperon
in nuclear matter, from hypernuclear spectroscopy and the lower bound of the mass of a neutron star; altogether they
determine the ratios of the hyperon/nucleon coupling constants, which correlate the relative strengths of the
$\sigma$ and $\omega$ mean-fields to the nucleon and hyperons:
\begin{eqnarray}
(B/A)=\chi_{\omega}V-\chi_{\sigma}S,
\end{eqnarray}
and where $V=(g_{\omega}/m_{\omega})^2\rho_0$ and $S=(M-M^\star)$ are the mean-field vector and scalar potentials,
respectively.

To determine the nuclear matter equation of state at finite temperature we use a set of model parameterizations
known as GM3\cite{glendenning92a}. With this set of parameters, we find for the nucleon effective mass, at
saturation density and for the compression modulus of nuclear matter,  results in the range of current experimental
values for these quantities as for instance the above cited ones. The effects of nonlinear scalar self-couplings are
studied by Koepf et. al.\cite{compression}. Though the ratios $\chi_i$ are known only qualitatively, studies have
shown however that QHD predictions of bulk static properties of neutron stars\cite{glendenning97} are very sensitive
to different choices for $\chi_i$.

Neutrinos, which are not included in the original approach for the GM3 parameters, are
 an indispensable agency for protoneutron star studies. Neutrinos are thus included in our formulation,
 at chemical equilibrium,
via the lepton fraction  ratio
\begin{eqnarray}
Y_L=\frac{\rho_e+\rho_\nu}{\rho_B},
\end{eqnarray}
where $\rho_e$, $\rho_{\nu}$ and $\rho_B$ represent, respectively, the electron, neutrino and baryon densities.
Explicitly, we extract the lepton fraction $Y_L$ from the study of the gravitational collapse of the core.

The values for the scalar ($\sigma$), vector ($\omega$) and isovector ($\varrho$) GM3 coupling constants,
$(g_i/m_i)^2$, with $i=\sigma,\omega,\varrho$,
 are shown in table \ref{const} together with
the hyperon/nucleon coupling constant ratios, and the lepton fraction, $Y_L$.
\begin{table}[ht]
\caption{ Parameters adopted in the model.\label{const}} \centerline{\footnotesize}
\begin{tabular}{l c c c c } \\
\hline  $(g_\sigma/m_\sigma)^2$ & $(g_\omega/m_\omega)^2$ & $(g_{\varrho}/m_{\varrho})^2$ & b ($\times 100$) & c
($\times 100$)
\\
 $fm^2$ & $fm^2$ & $fm^2$ &  &   \\  9.927 & 4.820 & 4.791 & 0.8659 & -0.2421 \\ \hline $\chi_{\sigma}$ & $\chi_{\omega}$ &
$\chi_{\varrho}$ & $Y_L$ & - \\ 0.6 & 0.568 & 1.0 & 0.4 & - \\ \hline \\
\end{tabular}
\end{table}

\begin{center}
\begin{figure}[htb]
\vspace*{10pt} 
\vspace*{1.4truein}             
\vspace*{10pt} \parbox[h]{4.5cm}{ \includegraphics{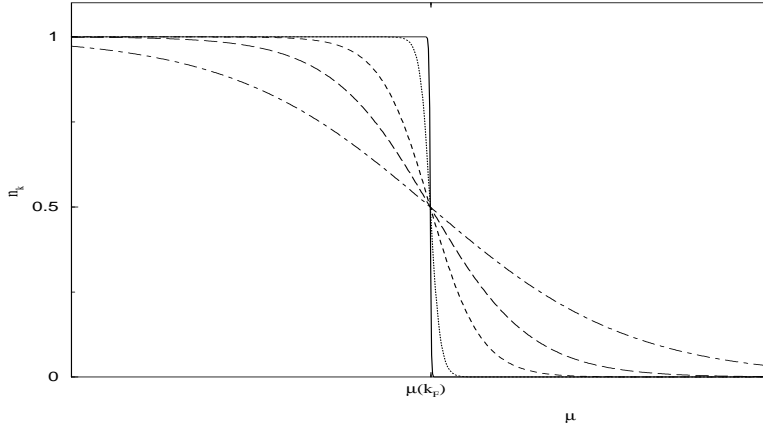}}
\vspace{20pt} \caption{Fermi distribution $n_k$ as a function of the baryon chemical potential for T=0MeV (solid
line), T=10MeV (dotted line), T=50MeV (dashed line), T=100MeV (long dashed line) and T = 200MeV (dot-dashed line).
\label{fermi}}
\end{figure}
\end{center}

\begin{figure}[htb]
\vspace*{10pt} 
\vspace*{1.4truein}             
\vspace*{10pt} \parbox[h]{4.5cm}{ \includegraphics{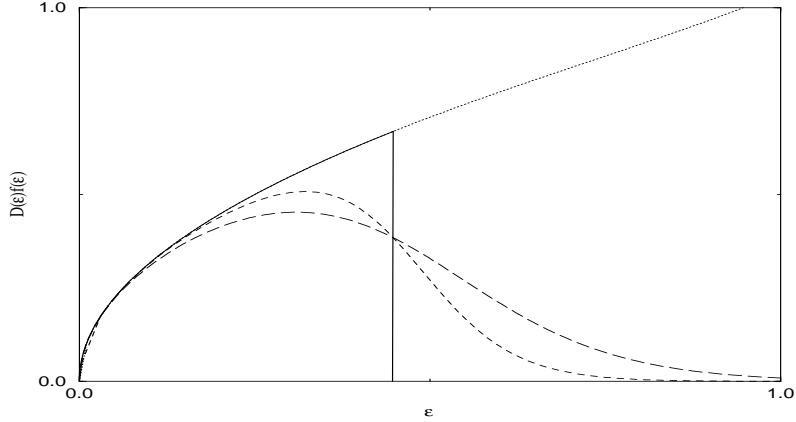}}
\vspace{20pt} \caption{Density of states as a function of baryon squared momentum for T=0MeV (solid line), T=50MeV
and T=100MeV \label{dns}}
\end{figure}

\begin{figure}[htb]
\vspace*{10pt} 
\vspace*{1.4truein}             
\vspace*{10pt} \parbox[h]{4.5cm}{ \includegraphics{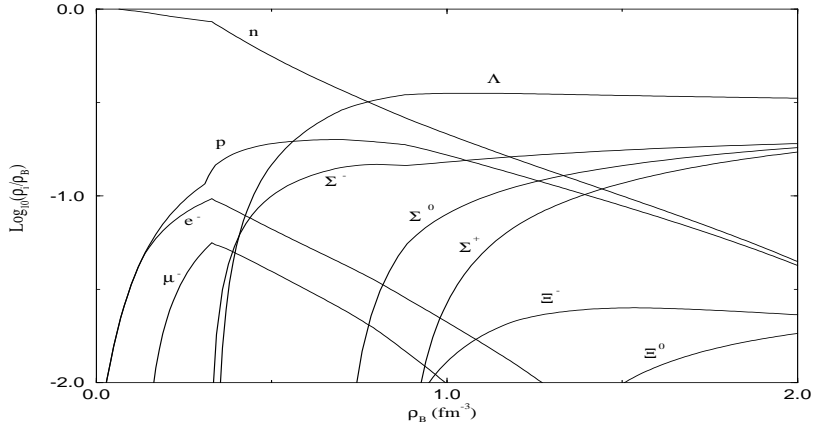}}
\vspace{20pt} \caption{Distribution of particles as a function of baryon density for T=10MeV, without neutrino
degrees of freedom. \label{free}}
\end{figure}

\begin{figure}[htb]
\vspace*{10pt} 
\vspace*{1.4truein}             
\vspace*{10pt} \parbox[h]{4.5cm}{ \includegraphics{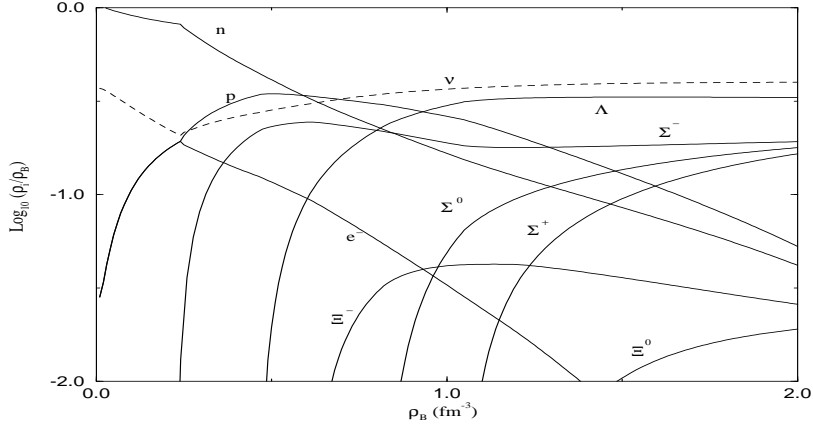}}
\vspace{20pt} \caption{As figure \ref{free}, including however neutrinos in the calculations. \label{trapped}}
\end{figure}

\begin{figure}[htb]
\vspace*{10pt} 
\vspace*{1.4truein}             
\vspace*{10pt} \parbox[h]{4.5cm}{ \includegraphics{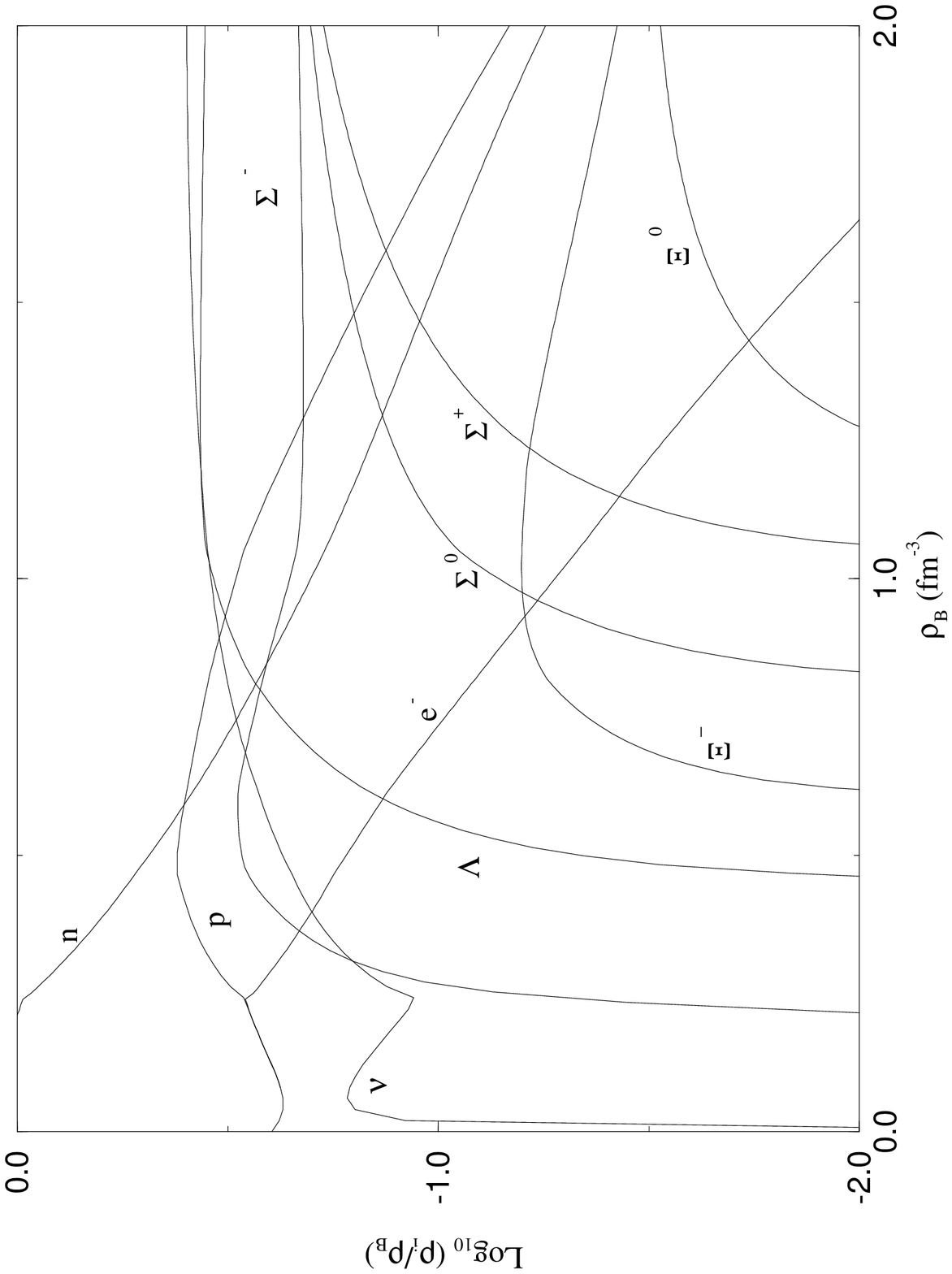}}
\vspace{20pt} \caption{As figure \ref{trapped}, however for T=50MeV. \label{trapt}}
\end{figure}

Applying standard technics from field theory and the mean-field approach, we obtain the equation of state for
nuclear matter as a parametric equation $p = p(\epsilon)$ which explicitly relates the energy density
\begin{eqnarray}
\epsilon &=& \frac{1}{2}m_\sigma^2\sigma^2 +\frac{1}{2}m_\omega^2\omega_0^2
+\frac{1}{2}m_\varrho^2\varrho_{03}^2+\frac{bM}{3}\sigma^3+\frac{c}{4}\sigma^4 \nonumber \\ &&+
\sum_B\frac{2J_B+1}{(2\pi)^3}\int_0^\infty \sqrt{k^2+{M_B^*}^2}[n(\mu_B,T)+{\bar n}(\mu_B,T)]d^3k \nonumber \\ &&+
\sum_\lambda\frac{2}{(2\pi)^3}\int_0^\infty \sqrt{k^2+{m_\lambda}^2}[n(\mu_\lambda,T)+{\bar n}(\mu_\lambda,T)]d^3k
\, ,  \label{eq1}
\end{eqnarray}
and pressure
\begin{eqnarray}
p &=& -\frac{1}{2}m_\sigma^2\sigma^2 +\frac{1}{2}m_\omega^2\omega_0^2
+\frac{1}{2}m_\varrho^2\varrho_{03}^2+\frac{bM}{3}\sigma^3+\frac{c}{4}\sigma^4 \nonumber \\ &&+ \frac{1}{3}
\sum_B\frac{2J_B+1}{(2\pi)^3}\int_0^\infty \frac{k^2}{\sqrt{k^2+{M_B^*}^2}} [n(\mu_B,T)+{\bar n}(\mu_B,T)]d^3k
\nonumber \\ &&+ \frac{1}{3} \sum_\lambda\frac{2}{(2\pi)^3} \int_0^\infty \frac{k^2}{\sqrt{k^2+{m_\lambda}^2}}
[n(\mu_\lambda,T)+{\bar n}(\mu_\lambda,T)]d^3k \, . \label{eq2}
\end{eqnarray}
In the expressions above we may identify, in the first row of both, the contributions of the scalar, vector and
isovector mesons, followed by the contributions of the Fermi gas of baryons and leptons; these contributions appear
as sums over the particle families, $B=n,p,\Sigma,\Xi,\Lambda$ and $\lambda=e,\mu,\nu$, and where $J_B$ is the
isospin baryon number. The quantities $n(\mu,T)$ and ${\bar n}(\mu,T)$ are the Fermi-Dirac distribution functions
for particle and anti-particles, at finite temperature:
\begin{eqnarray}
n(\mu_B,T) =[e^{\beta(\sqrt{\kappa^2+{M_B^*}^2}-\mu_B +\frac{g_{\omega}^2}{m_{\omega}^2}\rho_B
+\frac{g_\varrho^2}{m_\varrho^2}\rho_3)}+1]^{-1}, \nonumber \\ {\bar n}(\mu_B,T)
=[e^{\beta(\sqrt{\kappa^2+{M_B^*}^2}+\mu_B -\frac{g_{\omega}^2}{m_{\omega}^2}\rho_B
-\frac{g_\varrho^2}{m_\varrho^2}\rho_3)}+1]^{-1},
\end{eqnarray}
with $\beta = 1/\kappa T$ and the chemical potentials $\mu_B$ and $\mu_{\lambda}$ for baryons and leptons,
respectively (see Figs.\ref{fermi} and \ref{dns} where the Fermi distribution function and the density of states are
plotted as functions of chemical potential and the squared momentum $\varepsilon$, respectively).
In the numerical calculations with the BB model, we have considered hyperonic matter without (Fig.\ref{free}) and
with trapped neutrinos (Fig.\ref{trapped}), and at a higher temperature (T=50MeV in Fig.\ref{trapt}) in order to
evaluate  the effect of these quantities on protoneutron star properties.

\begin{figure}[htb]
\vspace*{10pt} 
\vspace*{1.4truein}             
\vspace*{10pt} \parbox[h]{4.5cm}{ \includegraphics{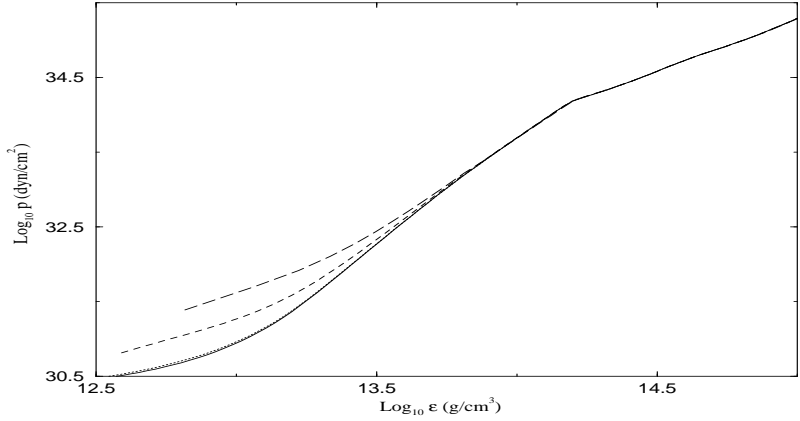}}
\vspace{20pt} \caption{Nuclear matter equation of state for T=0MeV (solid line), T=50MeV (dotted line) and T=100MeV
(dashed line). \label{eos}}
\end{figure}

\begin{figure}[htb]
\vspace*{10pt} 
\vspace*{1.4truein}             
\vspace*{10pt} \parbox[h]{4.5cm}{ \includegraphics{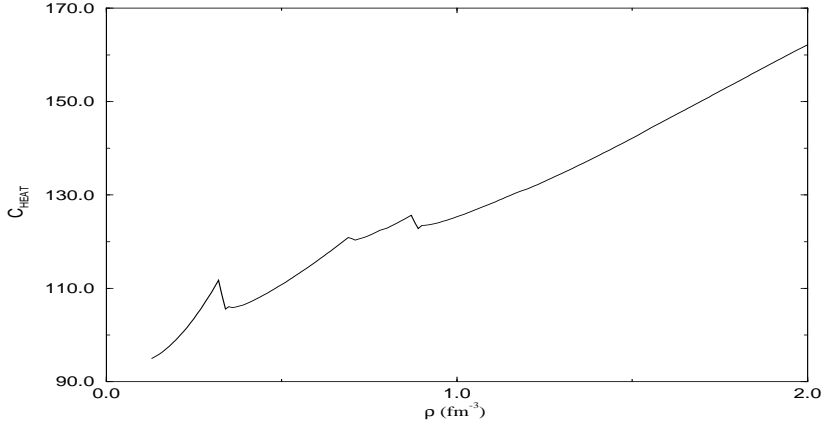}}
\vspace{20pt} \caption{Specific heat as a function of baryon density for T=10MeV. \label{calor}}
\end{figure}

A liquid-gas phase transition is predicted in various relativistic nuclear models such as QHD, BB and others, when
symmetric nuclear matter or pure neutron matter are described. However, following within the same models to describe
asymmetric nuclear matter, via the inclusion of the $\varrho$-meson, this phase transition disappears. Instead, the
energy used to generate this liquid-gas phase transition, leads now to hyperon production, resulting in a phase
transition of asymmetric nuclear matter into hyperonic matter. In such a phase transition, the system has to provide
an extra energy to convert nucleons into hyperons which makes the corresponding EOS stiffer. The provision of this
extra energy is not easily seen in this case in the EOS (Fig.\ref{eos}), as it is in the former Van der Waals
liquid-gas phase transition. It is however clearly visible in the curve for the specific heat, where each hyperon
generation formed introduces a discontinuity on its structure (Fig.\ref{calor}). This phase transition is similar to
the corresponding one with the formation of Cooper pairs in the supercondutivity phenomena in electron systems,
where discontinuities are also found instead of singularities, characterizing this type of behaviour as a continuous
phase transition.

\section{ Quark Matter}          

We have developed so far, nuclear matter studies disregarding the presence of quarks and gluons degrees of freedom.
However, the extreme conditions for the density in the interior of a protoneutron star may be favorable for the
occurrence of a hadron-quark phase transition, which may form a core of a quark-gluon plasma (QGP) in the inner
regions of the star.

We make use of the MIT bag model\cite{chodos} in order to describe the quark matter equation of state and the
hadron-quark phase transition. The MIT lagrangian is given as
\begin{eqnarray}
{\cal L}_{MIT}=[\frac{i}{2}({\bar\psi}\gamma^\mu\partial_\mu\psi-
(\partial_\mu{\bar\psi})\gamma^\mu\psi)-B]\Theta_V(x)-\frac{1}{2}{\bar\psi}\psi\Delta_s \, ,
\end{eqnarray}
(for details on the meaning of the symbols, see for instance\cite{MIT}). Following again standard  technical
procedures of quantum field theory, this lagrangian density generates the following equation of state for quark
matter ($p = p(\varepsilon)$):
\begin{eqnarray}
\epsilon=B+\sum_f\frac{\gamma_f}{2\pi^2}\int_0^\infty k^3dk [n_k(\mu)+{\bar n}_k(\mu)] \, ,
\end{eqnarray}
\begin{eqnarray}
p=-B+\sum_f\frac{1}{3}\frac{\gamma_f}{2\pi^2}\int_0^\infty k^3dk [n_k(\mu)+{\bar n}_k(\mu)] \, ,
\end{eqnarray}
with the degeneracy factor represented by $\gamma_f=2_{spin}\times 3_{colour}$ and the bag constant denoted by $B$.
The baryon density is given by
\begin{eqnarray}
\rho_B=\sum_f\frac{1}{3}\frac{\gamma_f}{2\pi^2}\int_0^\infty k^2dk [n_k(\mu)-{\bar n}_k(\mu)] \, .
\end{eqnarray}

The hadron-quark phase transition is determined via the Gibbs criteria\cite{Gibbs}, which require that, at constant

temperature, pressure and chemical potential of both hadron (H) and quark (Q) phases are related by
\begin{eqnarray}
P_Q=P_H,  \\  \nonumber \mu_Q=\mu_H  \, ,
\end{eqnarray}
for a conserved overall baryon number, i.e.
\begin{eqnarray}
N_Q/3 + N_H=constant \, ,
\end{eqnarray}
where $N_Q$ is the number of quarks and $N_H$ is the number of baryons in the system.

\begin{figure}[htbp]
\vspace*{10pt} 
\vspace*{1.4truein}             
\vspace*{10pt} \parbox[h]{4.5cm}{ \includegraphics{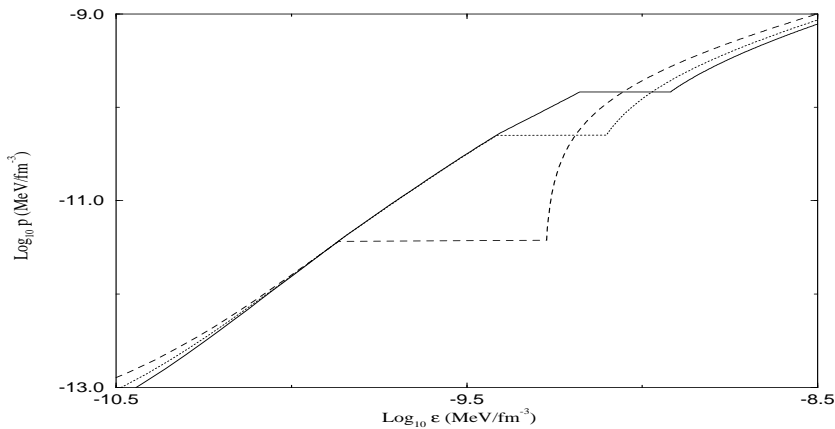}} \vspace{20pt} \caption{Equation of state with the hadron-quark phase transition for a bag constant
$B=100MeV/fm^3$(solid line), $B=131.2MeV/fm^3$ (dotted line) and $B=150MeV/fm^3$ (dashed line). \label{gfmfig1}}
\end{figure}

The resulting equation of state includes the presence of hadrons, with a phase of deconfined quark matter
represented in Fig.\ref{gfmfig1}, and where the plateau shown in the figure reflect the Gibbs condition of the phase
transition.

\section{ Stellar Structure}          

The protoneutron star configuration may be determined by combining the nuclear matter EOS with the TOV equations
from general relativity:

\begin{eqnarray}
4\pi r^2dp(r)=-\frac{M(r)dm(r)}{r^2}\left(1+\frac{p(r)}{\epsilon(r)}\right) \left(1+\frac{4\pi
r^3p(r)}{M(r)}\right)\left(1-\frac{2M(r)}{r}\right)^{-1},
\end{eqnarray}
and
\begin{eqnarray}
M(r)=4\pi\int^r_0 \epsilon (r^{\prime})r^{\prime \, 2} dr^{\prime}.\label{toveq}
\end{eqnarray}

The above equations describe the structure of a static, spherical and isotropic star wit the pressure $p(r)$ and the
energy density $\epsilon(r)$ reflecting the underlying nuclear model; $M(r)$ denotes the mass inside a sphere of
radius $r$. The TOV equations involve various constraints and boundary conditions: they must be evaluated for the
initial condition $\epsilon(0)=\epsilon_c$ (with $\epsilon_c$ denoting the central density) and $M(0)=0$ at $r=0$;
the radius R of the star is determined under the condition that on its surface the pressure vanishes, $p(r)|_{r = R}
= 0$.

The condition for chemical equilibrium for protoneutron stars are:
\begin{equation}
\mu_i = b_i\mu_n-q_i(\mu_{\ell}-\mu_{\nu_{\ell}})
\end{equation}
where $\mu_i$ and $\mu_{\ell}$ stand for the baryon and lepton chemical potentials, respectively;  $b_i$ is the
baryon number and the baryon and lepton electrical charges are represented by $q_i$.

The corresponding equations for baryon number and electric charge conservation are:
\begin{equation}
\rho_{baryonic} = \sum_B \frac{k_{F,B}^3}{3 \pi^2} \, ,
\end{equation}
and
\begin{equation}
\sum_B q_{e,B} \frac{k_{F,B}^3}{3 \pi^2} - \sum_{\ell} \frac{k_{F,\ell}^3}{3 \pi^2}=0 \, .
\end{equation}

\begin{figure}[htb]
\vspace*{10pt} 
\vspace*{1.4truein}             
\vspace*{10pt} \parbox[h]{4.5cm}{ \includegraphics{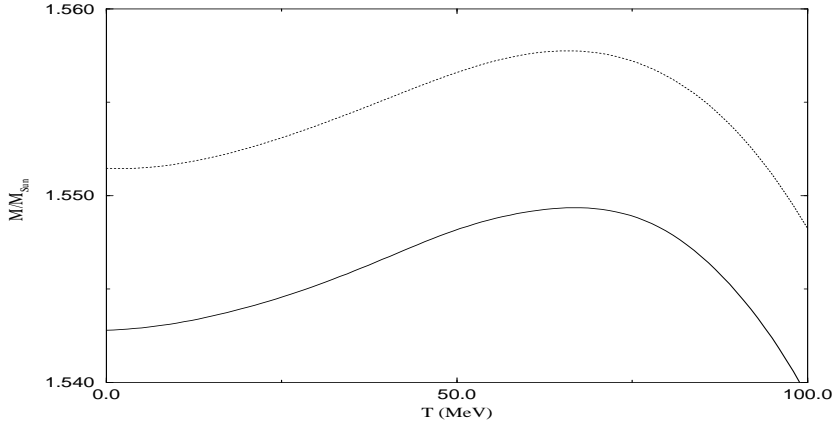}}

\vspace{20pt} \caption{Protoneutron star maximum mass as a function of temperature for $Y_l=0.4$ (solid line) and
$Y_L=0$ (dotted line). \label{pns1}}
\end{figure}

\begin{figure}[htb]
\vspace*{10pt} 
\vspace*{1.4truein}             
\vspace*{10pt} \parbox[h]{4.5cm}{ \includegraphics{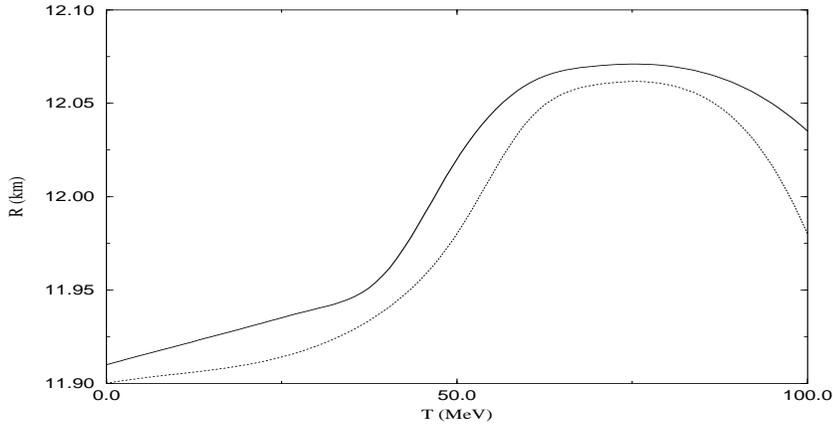}}
\vspace{20pt} \caption{Radii of protoneutron stars with maximum-mass as a function of temperature for trapped (solid
line) and for free (dotted line) neutrino matter, respectively. \label{pns2}}
\end{figure}

In the figures below, the bulk properties of the protoneutron star as a function of the temperature are shown.
Figs.\ref{pns1} and \ref{pns2} show respectively the maximum mass and its radius for hyperon matter including
neutrino trapping (solid curve) and the inverse situation, where neutrinos escape freely (dotted curve) These curves
indicate maximum masses respectively equal to $M_{lim}=1.547 M_{\odot}$ for the former case and $M_{lim}=1.557
M_{\odot}$ for the latter. Inspection of these figures indicates that for temperatures higher than 70 MeV
equilibrium is obtained for lower masses and radii. If $ T<70$ MeV the limiting mass increases with temperature.

The analysis of the maximum mass vs. temperature relationship is not trivial because of the complex structure of the
TOV equations and those describing the thermodynamical state of matter. In one hand, to the increase of temperature
it corresponds an increase of the mass of the protoneutron stars, a very well known result for the model used in
this calculation\cite{prakash97}; however, other models predict exactly the opposite behaviour. The novel feature we
have found without any doubt is an absolute limiting value for the protoneutron star mass, followed by a subsequent
decrease of this maximum value with the increasing of temperature. The existence of this maximum mass for
protoneutron stars, reached at $T\sim 70 MeV$ is a new result obtained in this study and we mention initially three
possible reasons to explain it: the very sensible balance of pressure, energy density and gravity in the TOV
equations; the modification of the lower limit on the energy density determined by thermal contributions, and the
complex behaviour of the scalar field at finite temperature. Here we briefly comment on the last two aspects and in
the following we present some analytical estimates for this behaviour.

As temperature increases, due to thermal contributions,  an increasingly minimal value for the energy density of
nuclear matter is progressively reached. Our results indicate that, when taking into account into the calculations
this characteristic behaviour of the thermal energy density, it implies sharper crusts for $T > 0$, as compared to
the corresponding results at zero temperature. As this effect starts to be significant around $T=50MeV$, it might be
a signal that the maximum protoneutron star mass starts to decrease.

A more microscopic and most interesting explanation for the results previously stressed, with striking consequences
for nuclear matter at high densities, can be found in the opposite behaviour of the meson scalar field at the high
and low density regimes. At low baryon densities, the scalar field does not change significantly as a function of
temperature until T reaches several tens of MeV; then, from this value on, it starts to increase. On the contrary,
for high baryon densities, the scalar meson field decreases as the temperature increases already beyond the first
few MeV. As the temperature increases, this behaviour becomes even more significant and thus might be basically
responsible for the protoneutron star maximum mass behaviour exhibited here.

The temperature dependence of the maximum star mass is supported by analytical estimates of the T-dependence of the
compression modulus and the equation of state of nuclear matter.

Concerning the compression modulus of nuclear matter,

\begin{eqnarray}
K(\rho_0, T) & = &  9 \rho_0^2 \left[ \frac{d p}{d \rho} \left. \right|_{\rho =\rho_0} \right] \nonumber \\ & = &  9
\rho_0^2 \left[ \frac{\partial p}{\partial \rho} \left. \right|_{\rho =\rho_0} + \frac{\partial p}{\partial T}
\left(\frac{\partial \rho}{\partial T}\right)^{-1} \left. \right|_{\rho =\rho_0}  \right] \, ,
\end{eqnarray}
we find for its leading dependence in T
\begin{equation}
K(\rho_0,T) = K(\rho_0, T=0)\left(1 + a T^2 \right) ,
\end{equation}
with $a = 0.0181$. This expression indicates that as temperature increases, the compression modulus of the
protoneutron star matter also increases, giving a stiffer equation of state, for a fixed density, in comparison to
the case $T=0$, with the immediate consequence that the stiffer the equation of state, the more massive becomes the
limiting star mass of a family of compact stars.

A second analytical argument involves the threshold equation for a given species of hadrons

\begin{equation}
q_B \mu_n - q_{e,B} \mu_e \geq g_{\omega B} \omega_0 + g_{\varrho B} \varrho_{0 3} I_{3 B} + (M_B - g_{\sigma B}
\sigma) \, ,
\end{equation}
where the isospin term $g_{\varrho B} \varrho_{0 3} I_{3 B}$ in the eigenvalue determines whether a species is
isospin-favored or unfavored. By using for the chemical potential\cite{woosley}
\begin{equation}
\mu_B \sim \mu_B(\!T\!=\!0) \left[1 - \frac{1}{3} \left(\frac{\pi \kappa T}{2 \mu_B(\!T\!=\!0\!)} \right)^2 \right]
\, ,
\end{equation}
with
\begin{equation}
\mu_B(T=0) = g_{\omega B} \omega_0 + g_{\varrho B} \varrho_{0 3} I_{3 B} + \sqrt{k_B^2 + (M_B - g_{\sigma B}
\sigma)^2} \, ;
\end{equation}
one finds (see also figures (3)-(5)), that at densities relevant for neutron stars the relative density distribution
of the hyperons, which contribute to the softening of the equation of state and thus to a reduction of the limiting
star mass, is shifted to higher values with increasing temperature. As a consequence, the limiting mass of the
protoneutron star will increase in the comparison with the corresponding value at $T=0$. Similar arguments were
recently used by Prakash\cite{prakash97} in his studies on compact stars a finite T.

Apart from these considerations, a more complete analysis should involve both TOV equations which allow, for an
appropriate discussion, only numerical solutions. On the other hand, still some insights may be obtained from the
T-dependence of the energy density $\epsilon(\kappa,T)$. Taking the Fourier transform at finite temperature,
\begin{eqnarray}
\epsilon(r,T) & =  & \frac{1}{(2 \pi)^{3/2}} \int \, \epsilon(\kappa,T) \, e^{i \kappa \cdot r} d^3 \kappa \nonumber
\\ & = &  \frac{1}{(2 \pi)^{3/2}} \int_0^{\infty}  d \kappa \, \kappa^2
\epsilon(\kappa,T) \int_0^{\pi} e^{i \kappa r cos(\theta)} sin(\theta) d \theta \nonumber \int_0^{2 \pi} d\phi \\ &
= &
 \frac{4 \pi}{(2 \pi)^{3/2}} \int_0^{\infty}  d \kappa \, \kappa^2 j_0(\kappa r)
\epsilon(\kappa,T) \, .
\end{eqnarray}
and combining this expression with equation (\ref{toveq}), we find the following equation for the protoneutron star
mass
\begin{equation}
M  =   \frac{4 \pi R}{(2 \pi)^{3/2}} \int_0^{\infty} \, d \kappa \, \epsilon(\kappa,T=0) \, j_1(\kappa R) + \frac{4
\pi R}{(2  \pi)^{3/2}} \int_0^{\infty} \, d \kappa \, \epsilon_{S}(\kappa,T) \, j_1(\kappa R) \, . \label{mass}
\end{equation}
Above, $R$ represents the protoneutron star radius and $\kappa$ is the Fermi momentum; furthermore, $j_i(\kappa R)$
with $i = 0,1$ represent Bessel functions of $i$-kind while $\epsilon_{S}(\kappa, T)$ denotes the temperature
corrections described in the appendix
\begin{eqnarray}
\epsilon_{S}(\kappa,T) & = &  \sum_{i=B} \frac{\gamma}{12}(k_BT)^2 [2k_{F,i}E_{F,i}+\frac{k_{F,i}^3}{E_{F,i}}]
\\ \nonumber &+& \frac{\gamma}{720}\pi^2(k_BT)^4[12\frac{k_{F,i}}{E_{F,i}}
-9\left(\frac{k_{F,i}}{E_{F,i}}\right)^3+3\left(\frac{k_{F,i}}{E_{F,i}} \right)^5] \, ,
\end{eqnarray}
with $k_{F,B} = \kappa$ (see equation (A.5)). Evidently, one identifies the first term in expression (\ref{mass}) as
the protoneutron star mass at T=0 and the second term as finite temperature corrections;  the Bessel function in
expression (\ref{mass}) modulates the kernel of the integral. As a crucial point, from the spectrum of the energy
density, low momenta $\kappa$ (with $\kappa R$ below the first zero of the Bessel function) dominate the integration
over $\kappa$ and add up coherently, resulting in an increase of the protoneutron star mass as a function of T.

We remark that our conjecture is supported both by a quite similar result in the non-relativistic limit obtained by
Chan\-dra\-sekhar\cite{chandra} in the description of the so called Chandrasekhar mass (see also Baron and
Cooper\-stein\cite{baron}) and from the studies recently performed by Woosley and Weaver on the birth of neutron
stars\cite{woosley}, who find for the temperature correction in the protoneutron star mass

\begin{equation}
M\approx M_{T=0}\left(1+\frac{\pi^2\kappa^2 T^2}{\epsilon_F^2}\right)
\end{equation}
with $\epsilon_F$ being the Fermi energy.

 The inclusion of quarks has a significant impact on the maximum mass of a neutron star:
hybrid stars may develop a QGP core, which may extend several kilometers, lowering its maximum mass due to the
stiffening of the EOS. For this reason, to develop a treatment compatible with the increasing of this value, we have
adopted here different parameterizations for the nuclear matter equation of state. At even higher central densities
one can find other classes of stars, for instance strange and quark stars, which are basically formed by deconfined
quark matter with a thin (1 km) crust of nuclear matter.

However, there is a very important condition which shall be taken into account when phase transitions are considered
in neutron stars. The Seidov  criterium\cite{seidov} establishes that whenever a density or/and an energy density
jump exists inside the star, consequence of a phase transition of any type, the resulting configuration is stable
only if satisfies the condition
\begin{equation}
\frac{\epsilon_2}{\epsilon_1}<\frac{3}{2}\left(1+\frac{p}{\epsilon_1}\right) \, ,
\end{equation}
where $p$ is the pressure at the interface between both phases, while $\epsilon_2$ and $\epsilon_1$ are respectively
the energy densities for the higher and lower density phases. It should be emphasized that this criterium is
applicable for fully relativistic star models.

The models described in this work generate hybrid proto-neutron stars which satisfy that condition at zero
temperature. However, objects hotter than 20 MeV are not able to satisfy the Seidov criterium and become unstable as
soon as they underwent a phase transition.

\section{ Summary and Conclusions}

In the framework of the Sommerfeld approximation, we have studied  the structure of protoneutron stars through an
effective many-body field theory at finite temperature. To the best of our knowledge, the approach used in this
work, including the Seidov criterium, was followed for the first time in the present investigation of this topic.
The Sommerfeld approximation permits a drastic simplification of the computational input and thus provides a rather
clear insight in the role of temperature on bulk static properties of these fascinating stellar objects.

The calculations were performed by using an extended version of the non-linear BB model, including the fundamental
baryon octet, $\sigma$, $\omega$ and $\varrho$ mesons, leptons degrees of freedom, trapped neutrinos,
--- introduced into the formalism at fixed lepton fractions, --- the equations for chemical equilibrium and charge
neutrality. Integrating the TOV equations we have obtained the maximum mass and the corresponding radius of the
protoneutron star as functions of the temperature.

The results indicate that to the increasing of temperature, it corresponds the increasing of the maximum mass and
radius of the star and, moreover, that the EOS of nuclear matter becomes stiffer. The causal limit is exactly
recovered in the EOS, even in the Sommerfeld approximation, where its recovering becomes more transparent. However,
at very high temperatures, the EOS acquires, compared to the T = 0 results,  a lower limit on the energy density
which prevents the increasing of the maximum mass of the protoneutron star; beyond a critical temperature, which we
have found equal to $T= 70MeV$, the maximum mass of protoneutron stars decreases.

We have analyzed the protoneutron star structure with and without neutrino trapping and the corresponding maximum
masses obtained are $1.557$ $M_{\odot}$ and $1.547$ $M_{\odot}$, respectively, for the set of parameters shown in
table \ref{const}. We have identified this value as the maximum protoneutron star mass which can be formed right
after the supernova event. With the same parameters in table \ref{const}, except for the choice of the
hyperon-nucleon coupling ratios\cite{glendenning97} $\chi_{\sigma}=\chi_{\omega}=\chi_{\varrho}=\sqrt{2/3}$, we have
obtained respectively $1.889$ $M_{\odot}$ and $1.868$ $M_{\odot}$. These last values are able to explain the mass of
the neutron star in the binary system Vela X-1\cite{pulsar}  and this example illustrates how these mass
determinations may guide us in the choice of adequate nuclear interaction coupling constants.

\begin{figure}[htb]
\vspace*{10pt} 
\vspace*{1.4truein}             
\vspace*{10pt} \parbox[h]{4.5cm}{ \includegraphics{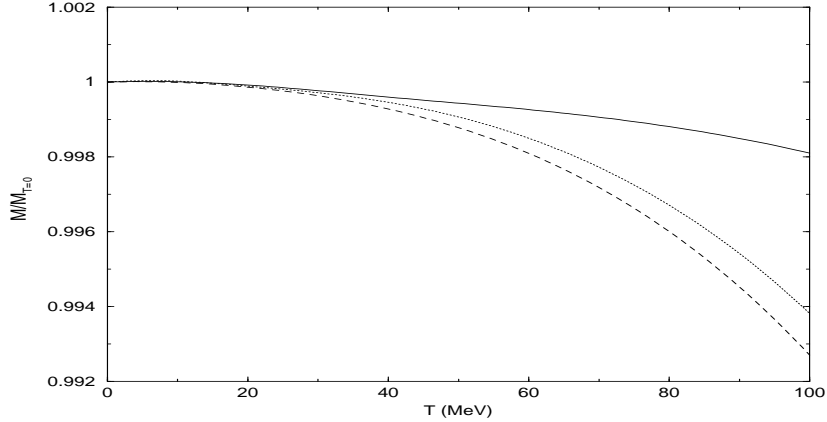}}
\vspace{20pt} \caption{Protoneutron star mass normalized by its mass at T=0, as function of temperature for
$M_{T=0}=1.8 M_\odot$ (solid line), $M_{T=0}=1.6 M_\odot$ (dotted line) and $M_{T=0}=1.35 M_\odot$ (dashed line).
\label{newf}}
\end{figure}

In figure \ref{newf} we exhibit the behaviour of the protoneutron star mass as the temperature increases (in order
to group three different stars in the same plot, we normalize the protoneutron star masses by their values at zero
temperature). Each of the three sets of stars are chosen to have the same total baryon number.

We also have seen that a hybrid protoneutron star, which shall obey the Seidov criterium, can only be formed at low
temperatures of the order of T=20MeV. Protoneutron stars hotter than that value shall first cool down as hyperon
stars and may form, subsequently, a quark core in its inner region.

The density dependence of the scalar meson field is of particular interest for dense hadronic systems in general, as
it touches upon the basic question how to organize effective field theories. Here two scenarios are obvious, which
lead to a drastically different dependence of the nonlinear scalar (and vector) meson field with density and
temperature. In the naive dimensional analysis, the scale parameter for natural nonlinear self-coupling of the
$\sigma$ and $\omega$ fields is the effective mass scale
\[
m_{eff} \sim \sqrt{f_{\pi} \Lambda} \sim (250 - 300 MeV) \, ,
\]
at saturation density, with $f_{\pi} \sim 93 MeV$ as the weak $\pi$-decay constant and $\Lambda \sim$ nucleon mass
as the scale for chiral symmetry breaking. From arguments based on Brown-Rho scaling and the onset of deconfinement
with the appearance of a quark-gluon plasma beyond some critical density $\rho_c \sim 4 \rho_0$, both $f_{\pi}$ and
$\Lambda$ and thus $m_{eff}$ decreases with increasing $\rho$ yielding
\[ m_{eff} \sim 0 \, for \, \rho > \rho_c \, .\]
Thus organizing effective field theory with the scale mass $\sqrt{f_{\pi} \Lambda}$ would completely change the role
of high order meson self-couplings --- in the limit $m_{eff} \to 0$ the corresponding lagrangian would develop an
essential singularity ---  and at the same time, used as a scale parameter for the temperature dependence
\[ \lambda \sim \frac{\kappa T}{m_{eff}} \, , \]
would invalidate the Sommerfeld approximation beyond a critical density.

However, an alternative scale for fermionic matter is set by the Fermi-momentum $k_{F}$ itself. At saturation
density the two scales coincide: at $\rho = \rho_0$
\[ m_{eff} \sim \sqrt{f_{\pi} \Lambda} \sim k_F \sim 270 \, MeV \, . \]
Opposite to $m_{eff}$, however, $k_F$ is stable, as it increases typically as $k_F \sim \rho^{1/3}$ with increasing
density. Such an organizational scheme of effective field theory would preserve both the dominance of the cubic and
quartic $\sigma$ (and $\omega$) self-couplings and with the expansion parameter
\[ \lambda \sim \frac{\kappa T}{k_F} \sim \frac{\kappa T}{(270 \, MeV) (\rho/\rho_0)^{1/3}} \, , \]

would even improve the convergence of the Sommerfeld approximation with increasing density.

Our investigations are just a first attempt towards the comprehension of the role of temperature  on physical
properties of neutron stars. Our main findings are substantiated  by calculations based on improved microscopic
models, which include as an example, in the spirit of the Brown-Rho scaling, a density dependence of the coupling
constants in the model. Investigations along these lines are in progress.

\appendix{\label{sommerf}}

The finite temperature studies involves many technical difficulties  through the Fermi integrals; in particular, the
computational effort increases dramatically. However, as from the integration of the full energy density only a
narrow range contributes significantly to observables, approximate schemes should provide at least semi-quantitative
results. The Sommerfeld approximation is a very successful approach first applied in condensed matter physics and
recently applied to the white dwarf problem\cite{ruffini}. We briefly describe, in this section, the Sommerfeld
approximation and discuss its validity when applied to nuclear matter problems. The main conception of the
approximation is the following: if a given function $H(\epsilon)$ of the energy density $\epsilon$ does not vary
rapidly in the energy range of the order of $\kappa T$ around $\mu$, the temperature dependence of the integral
$\int_{- \infty}^{\infty} H(\epsilon) n(\epsilon) d\epsilon$ should be given quite accurately by replacing
$H(\epsilon)$ by the first few terms in its Taylor expansion around $\epsilon = \mu$:
\begin{eqnarray}
\left. H(\epsilon)=\sum^\infty_{n=0}\frac{d^n}{d\epsilon^n} H(\epsilon)\right|_{\epsilon =
\mu}\frac{(\epsilon-\mu)^n}{n\!} \, .
\end{eqnarray}
The result is a series of the form
\begin{eqnarray}
\left. \int_{-\infty}^\infty H(\epsilon)n(\epsilon)d\epsilon = \int_{-\infty}^\mu H(\epsilon)d\epsilon +
\sum_{n=1}^\infty (\kappa T)^{2n}a_n\frac{d^{2n-1}}{d\epsilon^{2n-1}}H(\epsilon)\right|_{\epsilon=\mu} \, .
\end{eqnarray}
The coefficients $a_n$ are dimensionless constants of order ${\cal O}(1)$, and are determined according to
\begin{eqnarray}
a_n=\int_{-\infty}^\infty \frac{x^{2n}}{(2n)\!}\left(-\frac{d}{dx}\frac{1}{e^x+1}\right)dx.
\end{eqnarray}

The functions $H(\epsilon)$ of our approach have major variations on energy scales of the order of $\mu$ and thus in
general $(d/d \epsilon)^n H(\epsilon) |_{\epsilon = \mu}$ is of the order of $H(\mu)/ \mu^n$. Consequently,
successive terms in the Sommerfeld approximation are smaller by ${\cal O}((\kappa T)^2) \sim {\cal O} (10^{-4})$.
Consequently, in actual calculations only the first few terms must be retained. The final form of the integrals, up
to the third order, is then given as
\begin{eqnarray}
\int_{-\infty}^\infty H(\epsilon)f(\epsilon)d\epsilon &=& \int_{-\infty}^\mu H(\epsilon)d\epsilon +
\frac{\pi^2}{6}(\kappa T)^2 H^\prime(\mu) \nonumber \\ & + &  \frac{7\pi^4}{360}(\kappa T)^4
H^{\prime\prime\prime}(\mu) +{\cal O}(\frac{\kappa T}{\mu})^6 \, ,
\end{eqnarray}
where we recognize the zero temperature contribution corrected by terms which include explicitly the temperature.

We describe, in the following, the EOS of nuclear matter at finite temperature, extending the T = 0  energy density
$\epsilon_0$ and pressure $p_0$, by the leading T-dependent coefficients from the Sommerfeld approximation:
\vspace{-0.4cm}
\begin{eqnarray}
\epsilon &=& \epsilon_0 + \sum_{i=B,\ell} \frac{\gamma}{12}(k_BT)^2 [2k_{F,i}E_{F,i}+\frac{E_{F,i}^3}{k_{F,i}}] \\
\nonumber &+& \frac{\gamma}{720}\pi^2(k_BT)^4[12\frac{E_{F,i}}{k_{F,i}}
-9\left(\frac{E_{F,i}}{k_{F,i}}\right)^3+3\left(\frac{E_{F,i}}{k_{F,i}} \right)^5];
\end{eqnarray}
\begin{eqnarray}
p&=&p_0 + \sum_{i=B,\ell}\frac{\gamma}{12}(k_BT)^2 [\frac{4E_{F,i}^3}{k_{F,i}}-\frac{E_{F,i}^5}{k_{F,i}^3}] \\
\nonumber  &+& \frac{\gamma}{720}\pi^2(k_BT)^4[24\frac{E_{F,i}}{k_{F,i}}-39\left(\frac{E_{F,i}}{k_{F,i}}\right)^3
+9\left(\frac{E_{F,i}}{k_{F,i}}\right)^5+15\left(\frac{E_{F,i}}{k_{F,i}}\right)^7].
\end{eqnarray}

\begin{figure}[htb]
\vspace*{10pt} 
\vspace*{1.4truein}             
\vspace*{10pt} \parbox[h]{4.5cm}{ \includegraphics{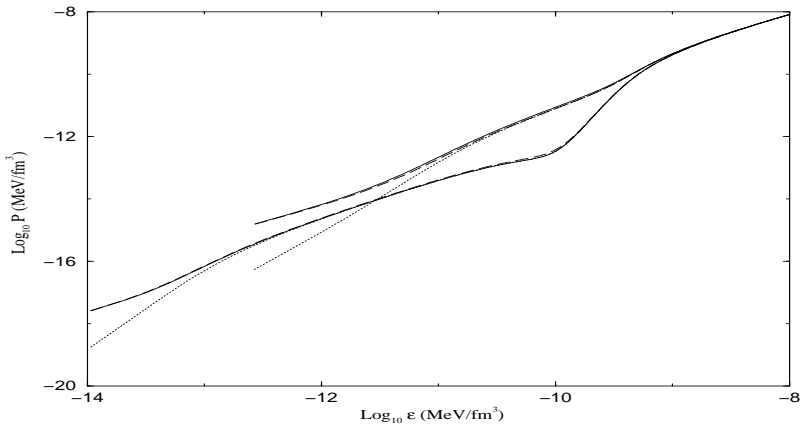}}
\vspace{20pt} \caption{EOS for the Walecka model for the temperatures T = 50 MeV and 10 MeV (upper and lower curves,
respectively). Solid lines: exact results; dashed and dotted lines: first and second order corrections of the
Sommerfeld approximation. \label{soc}}
\end{figure}

In previous appreciations, the Sommerfeld approximation has been used to describe systems in the high density and
low temperature ranges, $\frac{\mu}{k_BT}  >>1$, a condition which is very well satisfied in protoneutron stars,
with typically $ \frac{\mu}{k_BT} \sim [10 - 10^2]$; with this in mind, we feel confident about the adequacy of the
approach for the investigation here exhibited. However, to explore quantitatively the limiting range of validity of
the Sommerfeld approximation, we have performed calculations for less complex microscopic models and compared the
approximate results with numerically rigorous calculations (see figure (\ref{soc}).

We have used, in particular, the Sommerfeld approximation to describe the equation of state of the original QHD
model for pure neutron matter. Then, the equation of state was combined with the TOV equations, giving the results
shown in table \ref{tabelaa} for the exact calculations and the approximated ones using the Sommerfeld based
calculations. We have tested our results at the critical temperature for the liquid-gas phase transition in nuclear
matter (T=9.2MeV) and for higher values (T=50MeV), as typically found in protoneutron stars. For comparison, the
maximum mass predicted in the Walecka model at T = 0 is $2.57M_\odot$.

\begin{table}[h]
\begin{center}
\begin{tabular}{l c l c l c l} \hline \hline
& Temperature & Exact      & Sommerfeld & \\ \hline &      9.2MeV & 2.606      & 2.605      & \\ &       50MeV &
3.215      & 3.144      & \\   \hline\hline
\end{tabular}
\caption{Comparative results of the maximum protoneutron star mass. \label{tabelaa}}
\end{center}
\end{table}

The (nearly) quantitative results of the two approaches confirms the adequacy of the utilization of the Sommerfeld
approximation in the description of physical properties of nuclear matter at
finite temperature.

\indent

{\bf Acknowledgements}

This work was funded by CAPES and CNPq. The authors wish to thank Prof. M.A.C. Gusm\~ao for helpful discussions.

\end{document}